\documentclass[10pt,a4paper,twocolumn]{article}

\usepackage[T1]{fontenc}
\usepackage[utf8]{inputenc}
\usepackage[english]{babel}
\usepackage{amsmath,amssymb,bm}
\usepackage{graphicx}
\usepackage[a4paper,top=18mm,bottom=19mm,left=16mm,right=16mm,columnsep=6mm]{geometry}
\usepackage{mathptmx}
\usepackage{microtype}
\usepackage{url}

\graphicspath{{figures/}}
\title{Dynamical Modeling of Kinematically Decoupled Galaxies:\\Structural Components}
\author{V.~S.~Goradzhanov$^{1,2}$\thanks{E-mail: vgoradzanov@gmail.com},
D.~D.~Mishurin$^3$, O.~K.~Sil'chenko$^{1,2}$, and D.~F.~Gasymov$^1$\\[4pt]
\small $^1$Sternberg Astronomical Institute, Moscow State University, Moscow, 119234 Russia\\
\small $^2$Lomonosov Moscow State University, Moscow, 119991 Russia\\
\small $^3$School No.~2126 ``Perovo'', Moscow, 111524 Russia}
\date{Received January 30, 2025; revised April 23, 2026; accepted June 18, 2026}

\begin{document}
\maketitle

\begin{abstract}
We present the results of dynamical modeling of two galaxies with inconsistent kinematics -- PGC~35706 and LEDA~2220522 -- using data from the MaNGA integral-field spectroscopic survey and deep photometry from the DESI Legacy Imaging Surveys. The Schwarzschild orbit-superposition method in the axisymmetric approximation was employed to identify structural components based on the orbital circularity parameter $\lambda_z$. Three components have been identified in the galaxies: a spheroidal (bulge/halo) component and co-rotating and counter-rotating stellar disks. The spheroidal component dominates by mass in both galaxies: about 60\% in PGC~35706 and 51\% in LEDA~2220522; the contributions of the co-rotating and counter-rotating disks are 29\% and 9\% in PGC~35706 and 30\% and 18\% in LEDA~2220522, respectively. The kinematics of the ionized gas in both galaxies is consistent with the rotation of the counter-rotating disk, indicating that it formed from externally accreted gas with opposite angular momentum. The results support a scenario in which counter-rotating disks form through gas accretion from the intergalactic medium or through minor mergers with gas-rich satellites. This work demonstrates the effectiveness of the Schwarzschild method for studying the complex kinematic structure of galaxies and their formation history.

\medskip\noindent
\textbf{DOI:} \url{https://doi.org/10.1134/S1990341326600328}

\end{abstract}

\section{Introduction}

Studying the kinematics and dynamics of galaxies plays a key role in understanding the mechanisms of their formation and evolution. Of particular interest in this context are galaxies with inconsistent kinematics, in which different kinematic subsystems are observed, such as counter-rotating stellar disks, polar rings, or mismatched planes of rotation of stars and gas (Katkov et al., 2013, 2016). The study of such systems provides unique information about the gas-accretion and galaxy-merger processes that play a critical role in their evolution (Bournaud et al., 2005; Corsini, 2014).

A striking manifestation of the complex history of galaxy formation is the presence of counter-rotating stellar disks, first reliably detected in NGC~4550 (Rix et al., 1992; Rubin et al., 1992). Subsequent studies of individual objects (Coccato et al., 2011, 2013; Johnston et al., 2013; Katkov et al., 2024a) and galaxy samples from integral-field spectroscopic surveys (Bao et al., 2022; Bevacqua et al., 2022; Gasymov et al., 2025) have shown that counter-rotating disks are most often formed by external processes, such as gas accretion from the intergalactic medium or minor mergers with gas-rich satellites. In these scenarios, the newly formed stellar disk inherits the angular momentum of the accreted gas, which can be opposite to the angular momentum of the pre-existing disk. Comparing the ages and metallicities of the stellar populations of the two disks allows the sequence of events to be reconstructed and the dominant formation mechanism to be determined (Katkov et al., 2024b; Gasymov et al., 2025).

Dynamical modeling is a powerful tool for quantitatively studying the structural components of galaxies and their kinematics. In particular, the Schwarzschild orbit-superposition method (Schwarzschild, 1979) allows one to reconstruct the orbital distribution in a given gravitational potential, providing a physical separation of a galaxy into components (bulge, disk, halo) based on their orbital characteristics (van den Bosch et al., 2008; Zhu et al., 2018b). This method has been successfully applied to study orbital structures in large samples of galaxies from the CALIFA, MaNGA, and SAMI surveys (Zhu et al., 2018a; Jin et al., 2020; Santucci et al., 2022). The key parameter for classifying orbits is the circularity parameter $\lambda_z$, which allows rotating subsystems to be distinguished from subsystems supported by velocity dispersion (Zhu et al., 2018b).

Recently, the Schwarzschild method has been extended to include stellar-population information, leading to the development of population-orbit models (Zhu et al., 2020). This approach, tested on simulated data similar to MUSE observations (Bacon et al., 2010), has demonstrated the ability to reconstruct not only the orbit distribution but also the relationship between stellar kinematics and age, as well as the properties of different dynamical components. Furthermore, the first applications of nonparametric line-of-sight velocity distribution (LOSVD) descriptions and orbit-modeling methods to counter-rotating disks have appeared, overcoming the limitations of the Gauss-Hermite expansion of stellar LOSVDs in the presence of multiple kinematic components (Falc\'{o}n-Barroso and Martig, 2021; Gasymov and Katkov, 2022; Bao et al., 2024).

We begin a series of papers devoted to galaxies with inconsistent kinematics: their dynamical structure, the search for kinematically isolated subsystems, and the history of their formation and evolution. In this paper, we present the results of dynamical modeling of two galaxies from the sample of Gasymov et al. (2025): PGC~35706 with inconsistent kinematics and LEDA~2220522, which may have counter-rotating disks, based on data from the Mapping Nearby Galaxies at Apache Point Observatory (MaNGA) integral-field spectroscopic survey. Our goal is to identify and characterize their structural components, in particular their counter-rotating disks, using the Schwarzschild method implemented in the \texttt{Forstand} software package (Vasiliev and Valluri, 2020) within the AGAMA library (Vasiliev, 2019). We perform a multi-Gaussian expansion (MGE; Cappellari, 2002) photometric decomposition based on deep DESI images, construct a gravitational potential, and solve the inverse problem to find orbital weights that best reproduce the observational data. The resulting distributions of orbits in the phase space of radial distance and circularity allow us to estimate quantitatively the contributions of different dynamical components to the structure of the galaxies under study.

This paper is organized as follows. Section~2 describes the observational data and their processing. The photometric decomposition and construction of the gravitational potential are presented in Section~3. Section~4 describes in detail the process of dynamical modeling using the Schwarzschild method. The main modeling results and the analysis of the structural components are presented in Section~5. The results are discussed and summarized in Section~6.

\section{Data}

\subsection{MaNGA Integral-Field Spectroscopy}

This paper uses integral-field spectroscopy data for two galaxies obtained as part of the MaNGA survey (Westfall et al., 2019), which is part of the Sloan Digital Sky Survey-IV (SDSS-IV; Bundy et al., 2015). The survey provides a representative sample of more than 10,000 galaxies in the redshift range $0.01<z<0.15$, with a balanced stellar-mass distribution of $10^9$--$10^{11}\,M_\odot$ (Blanton et al., 2017).

The observations were made with the 2.5-m Sloan Foundation Telescope using the BOSS spectrographs (Smee et al., 2013), which provide simultaneous spectral coverage from 3600 to 10,000~\AA. The MaNGA data-reduction pipeline (DRP; Law et al., 2016) performs sky-background subtraction, flux calibration of the spectra, and creation of three-dimensional data cubes containing spatially resolved spectra. The wavelength-calibration accuracy is 5~km~s$^{-1}$ at a median spectral resolution of $R\sim2000$ (about 72~km~s$^{-1}$).

For our analysis, we require estimates of the kinematic parameters and stellar-population characteristics, so we perform full spectral fitting using the \textsc{NBursts} method (Chilingarian et al., 2007a,b) and a grid of simple stellar population (SSP) models computed with the E-MILES evolutionary-synthesis code (Vazdekis et al., 2016). Before fitting, the data-cube spaxels are binned using the Voronoi-binning method (Cappellari and Copin, 2003) to $S/N\sim20$.

When calculating the model spectrum during the $\chi^2$ minimization, the stellar-population spectrum is interpolated in the SSP-model grid by age $T_{\rm SSP}$ and metallicity $[\mathrm{Fe/H}]_{\rm SSP}$. The resulting spectrum is convolved with a parameterized stellar line-of-sight velocity distribution (LOSVD), described by a Gauss-Hermite function with coefficients $h_3$ and $h_4$. To reproduce the shape of the observed spectrum correctly, the model is multiplied by a 19th-degree polynomial continuum that compensates for differences in spectral shape caused by interstellar absorption and possible spectral-sensitivity calibration errors. The model also includes a set of bright emission lines (H$\gamma$, H$\beta$, [O\,III], [O\,I], [N\,II], H$\alpha$, [S\,III]), for which a single Gauss-Hermite LOSVD distinct from the stellar LOSVD is used. The emission lines are additive components whose weights (i.e., fluxes) are determined as a linear problem solved at each step of the nonlinear minimization cycle. The emission-line shapes and the resolution of the SSP models are pre-normalized to the wavelength-dependent instrumental resolution of the MaNGA data before the main minimization cycle. This simultaneous fitting of the stellar continuum and emission lines with independent kinematics is similar to \texttt{gandalf} (Sarzi et al., 2017) and recent versions of the \texttt{ppxf} code (Cappellari, 2017). As a result, we obtain two-dimensional maps of the kinematic parameters of the stellar and gaseous components: line-of-sight velocity $V$, velocity dispersion $\sigma$, and the Gauss-Hermite coefficients $h_3$ and $h_4$.

\subsection{Target Galaxies}

The galaxies were selected because they exhibit kinematic features indicating a complex dynamical structure. The sample comprises PGC~35706 (SDSS~J113356.96+511459.2) and LEDA~2220522 (SDSS~J105252.58+432542.2). Both galaxies are included in the final release of the MaNGA survey (DR17) and were observed using a 127-fiber integral-field unit, which provides good spatial coverage of their disk components.

\begin{figure*}[t]
\centering
\includegraphics[width=0.76\textwidth]{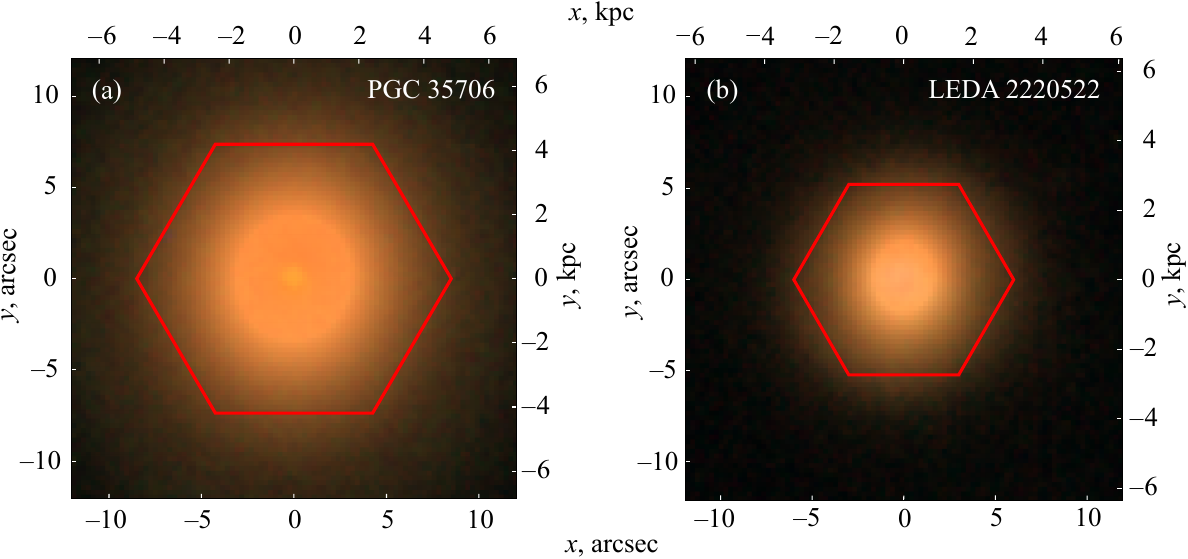}
\caption{Composite color images of galaxies PGC~35706 and LEDA~2220522, constructed from DESI survey images in the $g$, $r$, and $z$ filters. The red hexagon indicates the MaNGA field of view.}
\label{fig:desi}
\end{figure*}

\subsection{DESI Photometric Data}

To construct gravitational-potential models, along with the MaNGA spectral data, we use deep $z$-band images (Fig.~\ref{fig:desi}) obtained as part of the Dark Energy Spectroscopic Instrument survey (DESI; Dey et al., 2019). The high angular resolution and depth of the DESI images allow precise determination of the morphology of the studied galaxies and the parameters of their structural components, which is a critical input for dynamical modeling.

The choice of the $z$ band is driven by two main considerations. First, in this long-wavelength range the surface brightness best reflects the distribution of the old, mass-dominant stellar population and is less sensitive to local episodes of young star formation and to ionized-gas line emission, which is important for constructing an adequate mass-density model. Second, interstellar extinction in the $z$ band is minimal among the DESI optical channels (Schlafly and Finkbeiner, 2011), reducing systematic errors in surface-density reconstruction.

\section{Construction of the Gravitational Potential Profile}

The construction of accurate dynamical models of galaxies begins with photometric decomposition (Schwarzschild, 1979). Its goal is to decompose the observed surface-brightness distribution into structural components -- in our case, two-dimensional Gaussians -- and determine their parameters. These parameters form the basis for constructing the gravitational potential used in all subsequent dynamical modeling (Vasiliev, 2019).

\subsection{Photometric Decomposition and MGE}

The first step is to go from a two-dimensional surface-brightness map to a three-dimensional spatial mass density $\rho(x,y,z)$ for each component. Spheroidal components (bulges and halos) are often described by three-dimensional analytical models such as the Hernquist (Hernquist, 1990) or Navarro-Frenk-White (Navarro et al., 1997) profiles; the two-dimensional S\'{e}rsic (1968) profile, convenient for describing surface brightness, has the Einasto profile as a three-dimensional analogue. For disk components, exponential-disk models or the Miyamoto-Nagai (Miyamoto and Nagai, 1975) model are used.

For galaxies with complex morphologies, such as barred galaxies, simple analytical profiles may not be sufficiently flexible. In this paper, we use the multi-Gaussian expansion method (MGE; Cappellari, 2002), which is widely recommended for preparing dynamical models (Emsellem et al., 1994). In this approach, the two-dimensional surface-brightness profile is approximated by a sum of two-dimensional Gaussians. Each Gaussian is characterized by its total luminosity, dispersion along the major axis, and axial ratio.

\begin{figure*}[t]
\centering
\begin{minipage}{0.505\textwidth}
\includegraphics[trim=1cm 0.2cm 0.2cm 1.7cm,clip,width=\linewidth]{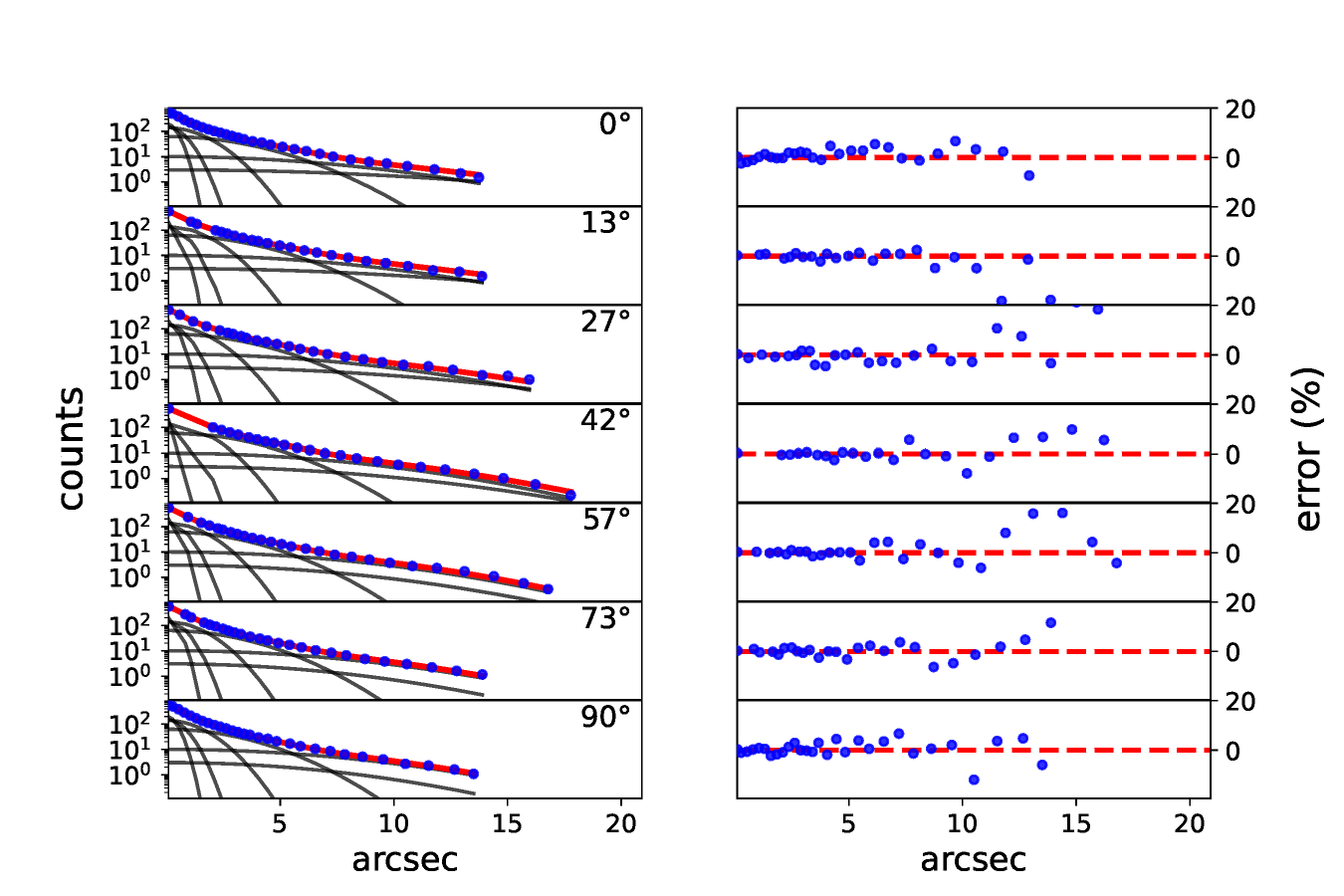}
\end{minipage}\hfill
\begin{minipage}{0.465\textwidth}
\includegraphics[trim=1cm 0cm 1.5cm 0cm,clip,width=\linewidth]{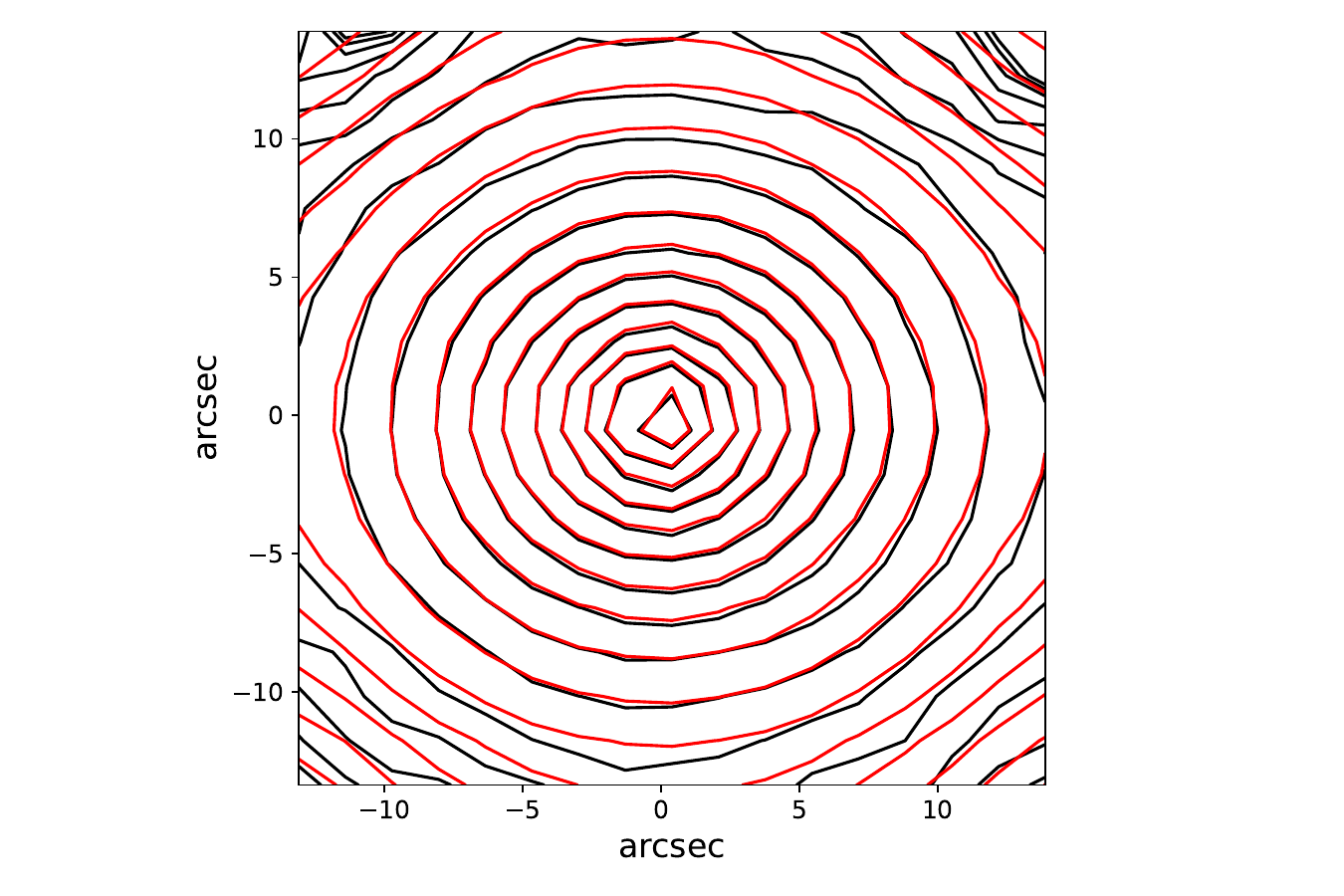}
\end{minipage}

\begin{minipage}{0.505\textwidth}
\includegraphics[trim=1cm 0.2cm 0.2cm 1.7cm,clip,width=\linewidth]{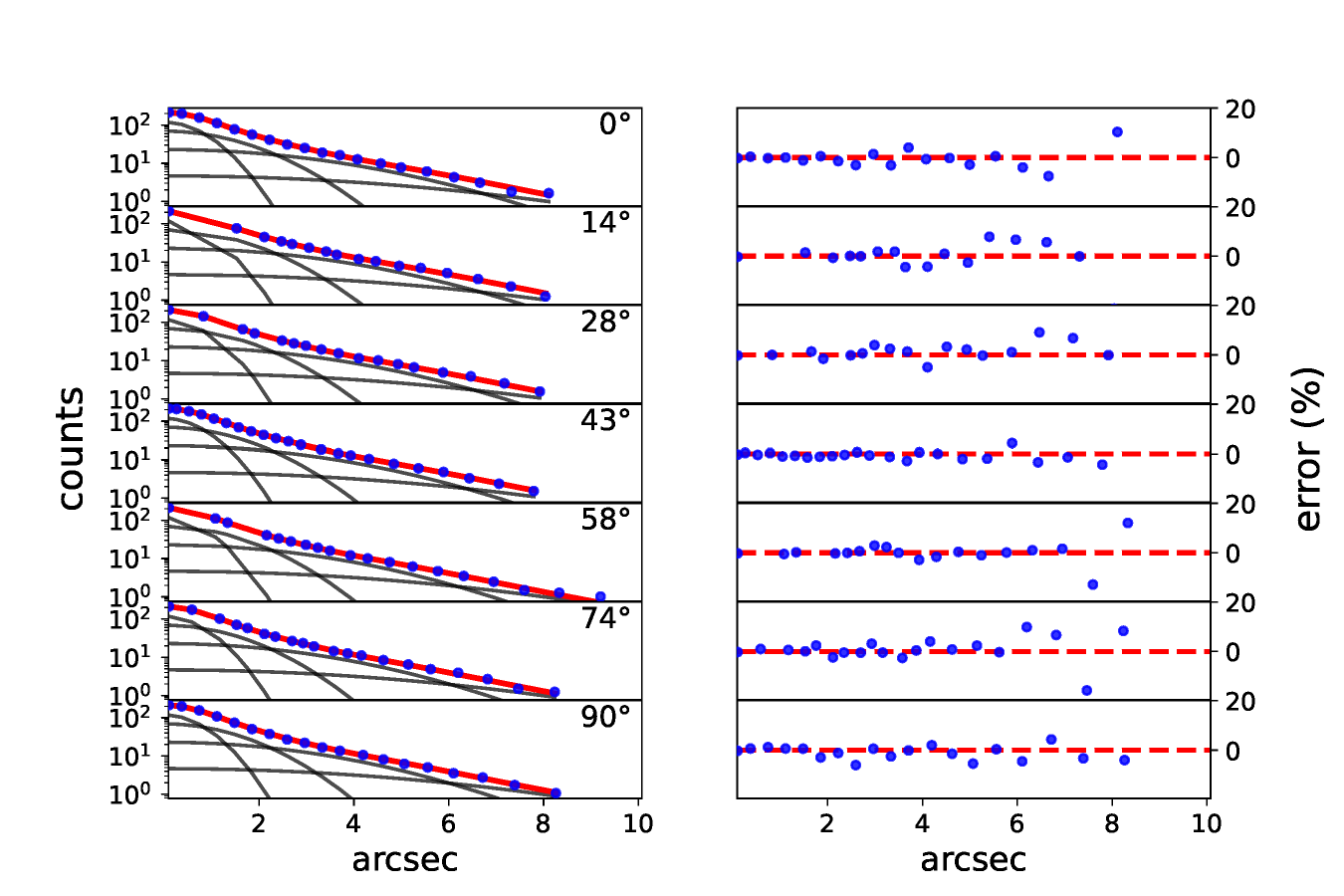}
\end{minipage}\hfill
\begin{minipage}{0.465\textwidth}
\includegraphics[trim=1cm 0cm 1.5cm 0cm,clip,width=\linewidth]{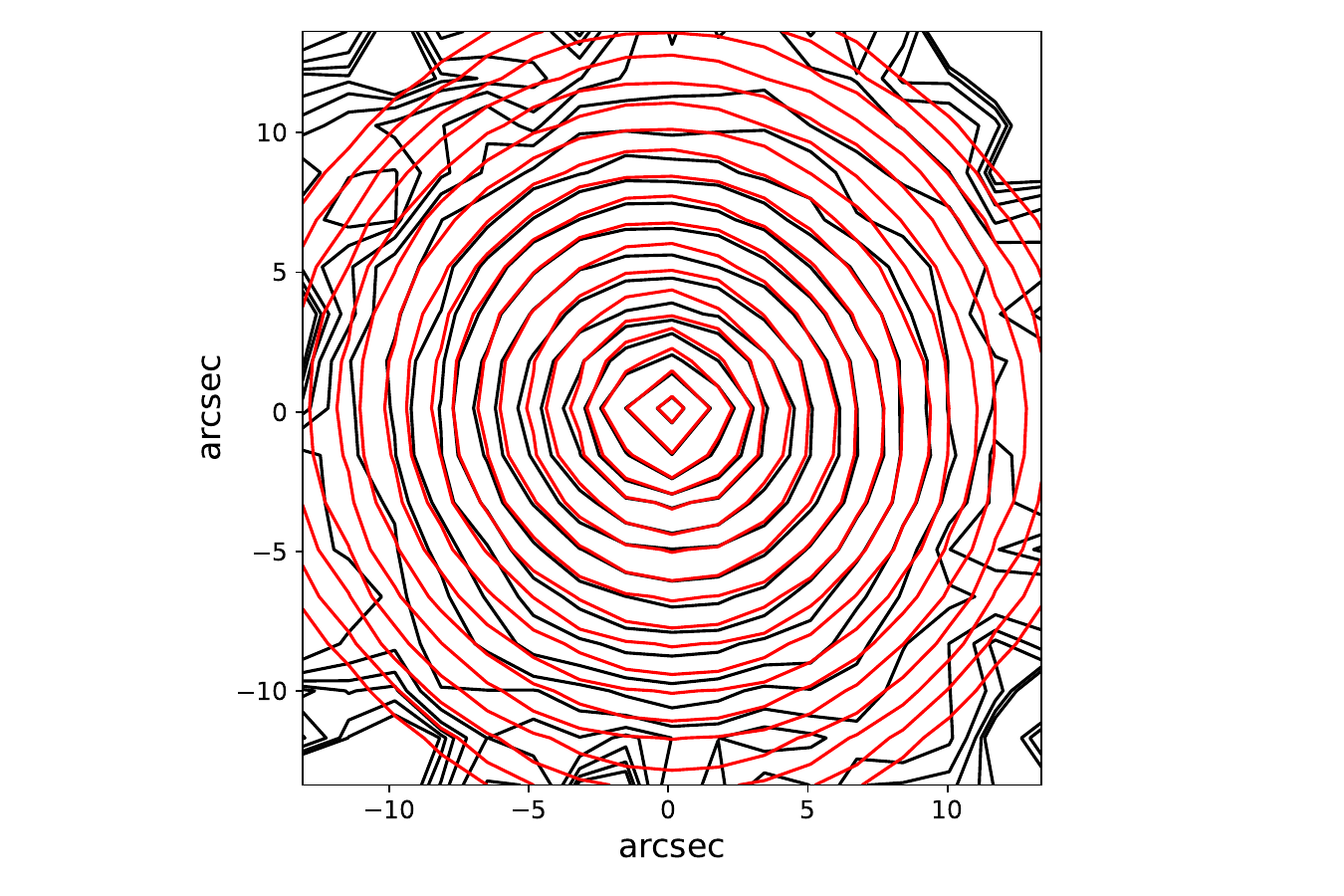}
\end{minipage}
\caption{Panel (a): comparison of the $z$-band photometric data for PGC~35706 (blue dots -- DESI photometry) with the best-fit MGE model represented as a sum of $N=6$ Gaussians (solid line). The individual convolved Gaussian components are also shown. From top to bottom, the profiles measured within $5^\circ$-wide sectors distributed over the angle between the major ($0^\circ$) and minor ($90^\circ$) axes are presented. Seven representative sectors are shown out of the $N_{\rm sec}=19$ sectors actually used in the model calculation. Panel (b): radial variation of the relative error along the profiles. Panel (c): galaxy isophote maps; the photometric data are marked in black and the MGE model in red. Panels (d)--(f): same as panels (a)--(c), but for LEDA~2220522.}
\label{fig:mge}
\end{figure*}

The MGE method provides fast and accurate calculation of the gravitational potential and its derivatives. The gravitational potential of an individual Gaussian is calculated analytically (Cappellari, 2002), and because an MGE is a linear combination of Gaussians, the overall potential is the sum of the analytical expressions for its components. This property follows from the shape of a Gaussian: the Poisson integral of a Gaussian density distribution reduces to a one-dimensional integral expressed through the error function, whereas no such reduction exists for other common surface-brightness parametrizations such as the S\'{e}rsic profile, for which the potential requires more complex numerical integration. The method also provides an analytical and smooth parametrization of the three-dimensional spatial density, which is needed for the subsequent solution of the Poisson equation.

The transition from two-dimensional Gaussians describing the observed surface-brightness distribution to a three-dimensional spatial-density distribution is made under the assumptions of axial symmetry and a specified inclination angle $i$. For each two-dimensional Gaussian with major-axis dispersion $\sigma_j$ and observed axial ratio $q'_j$, the corresponding three-dimensional Gaussian with the same $\sigma_j$ and intrinsic axial ratio
\begin{equation}
q_j=\sqrt{\frac{{q'_j}^{,2}-\cos^2 i}{\sin^2 i}}
\end{equation}
is reconstructed while preserving the full luminosity of the component (Emsellem et al., 1994; Cappellari, 2002).

The photometric profiles were analyzed as follows.
\begin{enumerate}
\item An initial estimate of the galaxy center and ellipticity was made.
\item Photometric data were sampled from the observed pixels with a logarithmic step in sectors distributed uniformly in angle while accounting for the galaxy ellipticity. To account for the spatial resolution, the profiles were convolved with the point-spread function (PSF). We used the PSF models provided by the DESI Legacy Imaging Surveys pipeline for the corresponding $z$-band images (Dey et al., 2019).
\item For each sector, the Gaussian dispersions along the major axis and, where necessary, the axial ratios were fitted. A fully linear algorithm for 200 Gaussians was used first, after which the model parameters were optimized with the nonlinear Levenberg-Marquardt algorithm and converged rapidly. Because this algorithm is sensitive to the initial approximation, the optimization was repeated with perturbed initial values, and the model with the minimum $\chi^2$ was selected from the resulting family of solutions. The stability of the solution was also verified visually by comparing the model with the data in all angular sectors.
\end{enumerate}
After optimization, the final profile model for each galaxy comprised six Gaussians with nonzero luminosities (Fig.~\ref{fig:mge}).

DESI error maps were not explicitly considered in the fitting. We minimized
\begin{equation}
\chi^2_{\rm ph}=\sum_i\left(\frac{I_i^{\rm obs}-I_i^{\rm mod}}{I_i^{\rm obs}}\right)^2,
\end{equation}
where $I_i^{\rm obs}$ is the observed surface brightness and $I_i^{\rm mod}$ is the PSF-convolved MGE model; the sum is over all selected points in all angular sectors. This logarithmically uniform weighting gives balanced weight to both the bright central regions and the extended low-surface-brightness parts of the galaxy and is the default implementation in the \texttt{mgefit} package (Cappellari, 2002).

\subsection{Construction of the Gravitational Potential}

The self-consistent gravitational potential $\Phi$ is then calculated by solving the Poisson equation
\begin{equation}
\nabla^2\Phi(\bm{x})=4\pi G\rho(\bm{x}),
\end{equation}
where $\rho(\bm{x})$ is the total spatial mass density obtained from the photometric decomposition.

At this stage, we used the AGAMA library (Vasiliev, 2019), which provides highly efficient solvers. For spheroidal components a spherical-harmonic expansion is used, whereas an azimuthal-harmonic expansion is preferred for highly flattened disk components. Both methods use spline interpolation on a precalculated grid, ensuring accurate calculation of the potential and its derivatives at any point in space. When constructing the potential with the AGAMA multipole method, we used a spherical-harmonic expansion with $l_{\max}=32$ for axisymmetric geometry ($m_{\max}=0$) on a radial grid of 40 nodes. The high $l_{\max}$ was chosen to reproduce adequately the potential generated by the flattened and inclined disk components of the galaxies. Thus, combining the MGE photometric decomposition with efficient Poisson-equation solvers from AGAMA yields a reliable and physically motivated gravitational-potential model for subsequent Schwarzschild modeling.

\section{Dynamical Modeling}

To study the structural components of the galaxies, we used the \texttt{Forstand} software package (Vasiliev and Valluri, 2020), part of the public AGAMA library (Vasiliev, 2018). The models of PGC~35706 and LEDA~2220522 were constructed under the assumption of an axisymmetric mass distribution.

\subsection{Gravitational Potential}

Constructing a Schwarzschild model requires specifying the gravitational potential formed by stars and dark matter. The dark-halo potential is specified by a modified Navarro-Frenk-White profile with an exponential cutoff at $r_{\rm cutoff}=100r_{\rm scale}$ to eliminate the mass divergence:
\begin{align}
\rho(\widetilde r)={}&\rho_0\left(\frac{\widetilde r}{r_{\rm scale}}\right)^{-1}
\left[1+\left(\frac{\widetilde r}{r_{\rm scale}}\right)\right]^{-2}\nonumber\\
&\times\exp\left[-\left(\frac{\widetilde r}{r_{\rm cutoff}}\right)^2\right],
\end{align}
where $\rho_0=(v_{\rm circ}/r_{\rm scale}/0.465)^2(4\pi G)^{-1}$, and $v_{\rm circ}$ and $r_{\rm scale}$ are free model parameters: the peak circular velocity and the characteristic halo radius.

The stellar-component potential obtained in Section~3 is added to the dark-halo potential to yield the total potential. The model includes an additional free parameter, the mass-to-light ratio $\Upsilon$, common to the entire stellar MGE model within the MaNGA field of view; no spatial variation in $\Upsilon$ is introduced. This parameter allows the depth of the gravitational potential to be adjusted without changing its shape and can compensate for possible mass underestimation associated with the photometric data and the dark-halo model.

\begin{figure*}[t]
\centering
\begin{minipage}{0.47\textwidth}
\includegraphics[width=\linewidth]{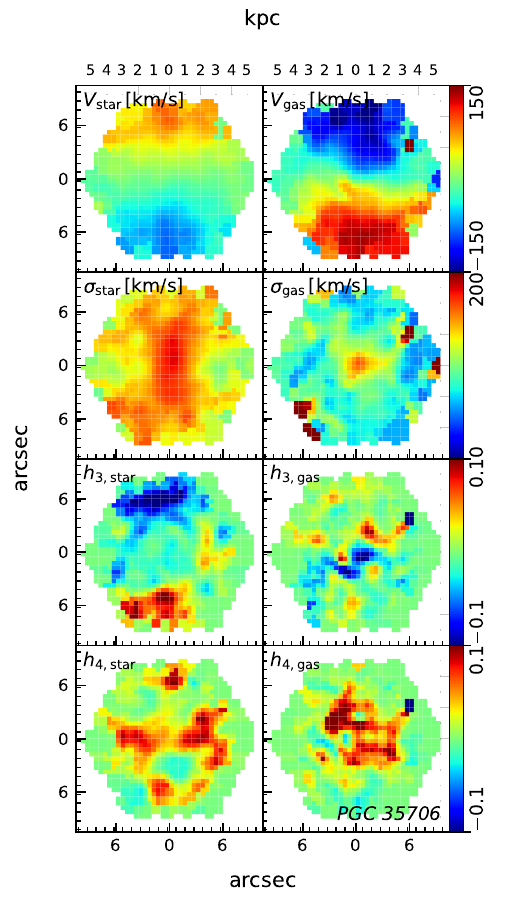}
\end{minipage}\hfill
\begin{minipage}{0.47\textwidth}
\includegraphics[width=\linewidth]{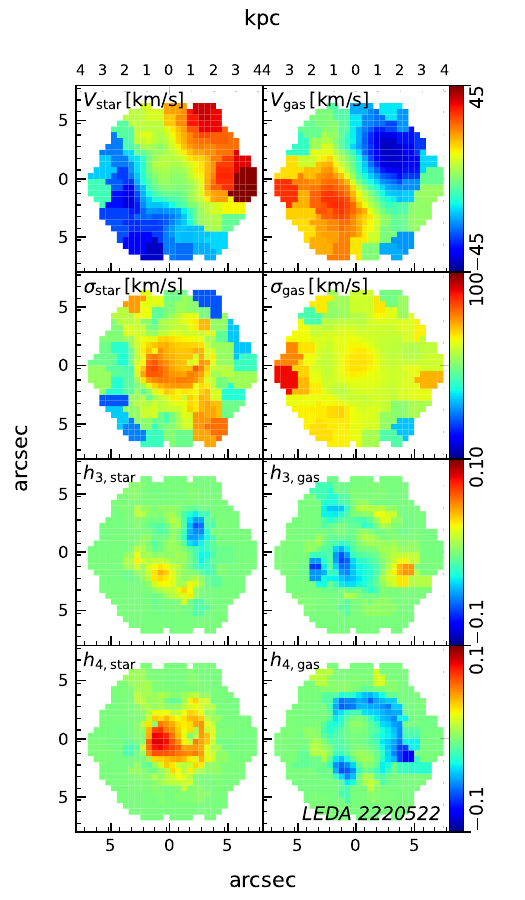}
\end{minipage}
\caption{Observed line-of-sight velocity $V$ (row a), velocity dispersion (row b), and Gauss-Hermite $h_3$ (row c) and $h_4$ (row d) fields for stars (columns 1 and 3) and ionized gas (columns 2 and 4), from MaNGA data. PGC~35706 is shown in columns 1--2 and LEDA~2220522 in columns 3--4.}
\label{fig:manga}
\end{figure*}

\subsection{Model Parameters}

The kinematic parameters are the parameters of the line-of-sight velocity distribution (LOSVD) obtained by fitting the MaNGA spectra as described above (Chilingarian et al., 2007a,b; Fig.~\ref{fig:manga}). The spatial resolution of the data is taken into account through convolution with the instrumental PSF.

The geometric parameters include the orientation angles of the galaxy relative to the line of sight. For axisymmetric systems, specifying the inclination angle is sufficient, whereas triaxial systems require three Euler angles. Rotating structures such as bars or spiral arms require an additional pattern-speed parameter $\Omega$, which determines the rotation of the reference frame in which the potential is stationary. Here we adopt an axisymmetric model.

We attempted to refine three input parameters: $v_{\rm circ}$, $r_{\rm scale}$, and $\Upsilon$. For each pair $(v_{\rm circ},r_{\rm scale})$, we constructed a gravitational potential and a corresponding orbit library. Changing $\Upsilon$ does not require recalculating the potential; its effect is included by scaling the velocities in the orbit library by $\sqrt{\Upsilon}$. Using the \texttt{AdaMet} algorithm (Cappellari et al., 2013), we found optimal values of $v_{\rm circ}$, $r_{\rm scale}$, and $\Upsilon$ from models with $N_{\rm orb}=10,000$ orbits. We then constructed a dynamical model with $N_{\rm orb}=40,000$ and the optimal circular velocity and halo radius. The mass-to-light ratio was further refined within the Schwarzschild simulations.

The quality of the model at each sample point in parameter space is estimated using
\begin{equation}
\chi^2_{\rm tot}=\chi^2_{\rm ph}+\sum_p\sum_{k\in\{V,\sigma\}}
\left(\frac{X^{\rm obs}_{k,p}-X^{\rm mod}_{k,p}}{\sigma^{\rm obs}_{k,p}}\right)^2,
\end{equation}
where $\chi^2_{\rm ph}$ combines the penalties for deviations of the model masses in the cells of the three-dimensional density grid and in the Voronoi apertures from the values prescribed by the MGE model, with a relative tolerance of 0.01. In the second term, $p$ indexes the Voronoi apertures; $X^{\rm obs}_{k,p}$ and $X^{\rm mod}_{k,p}$ are the observed and model stellar line-of-sight velocity and velocity dispersion in an aperture; and $\sigma^{\rm obs}_{k,p}$ is the corresponding uncertainty obtained by fitting the spectra with \textsc{NBursts} (Chilingarian et al., 2007a,b). The minimization has three nested levels. At the inner level, for fixed potential parameters and $\Upsilon$, the quadratic-programming problem for the orbital weights is solved (Vasiliev and Valluri, 2020). At the intermediate level, $\Upsilon$ is scanned on a one-dimensional logarithmic grid. At the outer level, the dark-halo parameters $(v_{\rm circ},r_{\rm scale})$ are selected with \texttt{AdaMet}.

For disk subsystems, the orbit library is initialized with the Jeans modeling method (Cappellari, 2008), with free parameters including the anisotropy
\begin{equation}
\beta=1-\frac{\sigma_\theta^2+\sigma_\phi^2}{2\sigma_r^2},
\end{equation}
where $\sigma_r(r)$, $\sigma_\theta(r)$, and $\sigma_\phi(r)$ are the radial, meridional, and azimuthal velocity dispersions, and the epicyclic frequency
\begin{equation}
\kappa(R)=\left.\sqrt{\frac{d^2\Phi_{\rm eff}}{dR^2}}\right|_{R_0}
=\sqrt{R\frac{d\Omega^2}{dR}+4\Omega^2},
\end{equation}
where $R$ is the distance from the galactic center in the disk plane, $\Phi(R)$ is the gravitational potential, and $\Omega(R)=V_c(R)/R$ is the angular velocity of circular motion at radius $R$. For spherical subsystems, the orbit library is initialized using the Eddington distribution function. Test particles are selected from the resulting distributions, and their trajectories are integrated for a time substantially longer than the orbital period.

The best input parameters are found with \texttt{AdaMet} (Cappellari et al., 2013), an adaptive implementation of the Metropolis algorithm (Haario et al., 2001). Its key feature is automatic adjustment of the covariance matrix of a multivariate Gaussian proposal distribution during Markov-chain Monte Carlo sampling. Unlike the classical Metropolis algorithm, in which this distribution is fixed, \texttt{AdaMet} periodically recalculates the covariance matrix from the accumulated chain history, ensuring convergence to the covariance matrix of the posterior distribution. This approach substantially accelerates convergence when parameters are strongly correlated and the posterior contains narrow, elongated regions. Although adaptation violates the Markov property of the chain, the algorithm preserves the correct ergodic properties and remains suitable for statistical inference. To accelerate the calculations further and reduce the risk of convergence to local minima, the initial position of the chain is determined with the global DIRECT optimization algorithm, eliminating a lengthy burn-in phase.

\begin{table*}[t]
\centering
\caption{Model parameters. $\Upsilon$ is the mass-to-light ratio in the $z$ band in solar units; $r_{\rm scale}$ is the characteristic dark-halo radius; $v_{\rm circ}$ is the peak circular velocity in the halo, reached outside the MaNGA field of view; $i$ is the inclination of the galaxy to the line of sight relative to the polar axis; $D$ is the distance from Gasymov et al. (2025), adopting WMAP9 cosmology ($H_0=69.3$~km~s$^{-1}$~Mpc$^{-1}$, $\Omega_0=0.2865$; Hinshaw et al., 2013); and $M_{*,\rm tot}$ is the total stellar mass of the model extrapolated beyond the field of view. The higher $v_{\rm circ}$ of LEDA~2220522 despite its lower stellar mass reflects the different dark-halo structures of the two galaxies.}
\begin{tabular}{lrrrrrr}
\hline
Galaxy & $\Upsilon$ & $r_{\rm scale}$ (kpc) & $v_{\rm circ}$ (km s$^{-1}$) & $i$ (deg) & $D$ (Mpc) & $M_{*,\rm tot}$ ($10^{10}M_\odot$)\\
\hline
PGC~35706 & 12.4 & 61 & 154 & 42 & 117 & 6.2\\
LEDA~2220522 & 3.6 & 69 & 324 & 27 & 109 & 0.55\\
\hline
\end{tabular}
\label{tab:model}
\end{table*}

\begin{figure*}[t]
\centering
\includegraphics[width=0.98\textwidth]{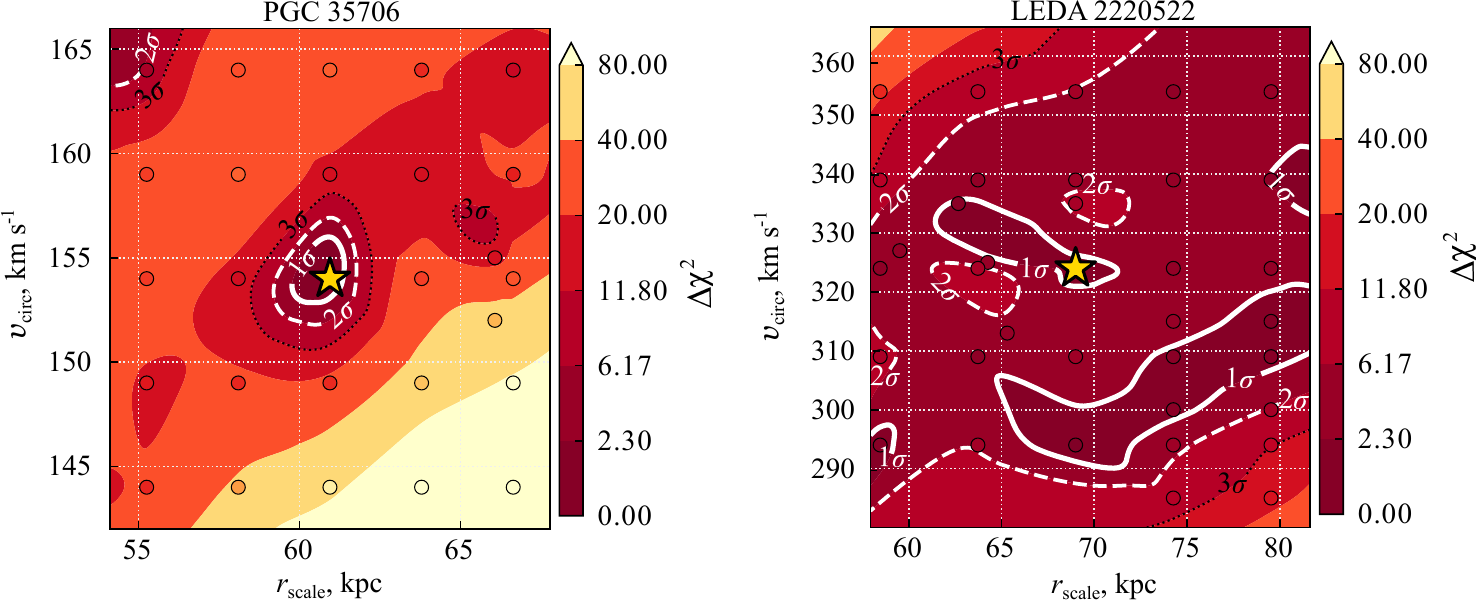}
\caption{Maps of $\Delta\chi^2=\chi^2-\chi^2_{\min}$ in the plane of dark-halo parameters $(r_{\rm scale},v_{\rm circ})$ at the optimal mass-to-light ratio $\Upsilon$. The asterisk marks the global minimum and the circles are the grid nodes tested by the \texttt{AdaMet} algorithm. Solid white, dashed white, and dotted black lines denote $\Delta\chi^2=2.30$, 6.17, and 11.83, respectively (1$\sigma$, 2$\sigma$, and 3$\sigma$ for two free parameters).}
\label{fig:chi2map}
\end{figure*}

Table~\ref{tab:model} gives the best-fit parameters. Figure~\ref{fig:chi2map} illustrates how well the data constrain the dark-halo parameters by showing maps of $\Delta\chi^2\equiv\chi^2-\chi^2_{\min}$ in the $(r_{\rm scale},v_{\rm circ})$ plane at the optimal $\Upsilon$. The contours at $\Delta\chi^2=2.30$, 6.17, and 11.83 correspond to the 1$\sigma$, 2$\sigma$, and 3$\sigma$ confidence regions for two free parameters (Avni, 1976). For PGC~35706, the minimum is well localized, indicating reliable reconstruction of the halo parameters within the MaNGA field of view. For LEDA~2220522, the 1$\sigma$ region is noticeably elongated, reflecting a moderate degeneracy between $r_{\rm scale}$ and $v_{\rm circ}$ expected from the limited radial coverage of the kinematic data; nevertheless, the minimum is stable and well separated from the 3$\sigma$ boundary.

\begin{figure*}[t]
\centering
\includegraphics[width=0.96\textwidth]{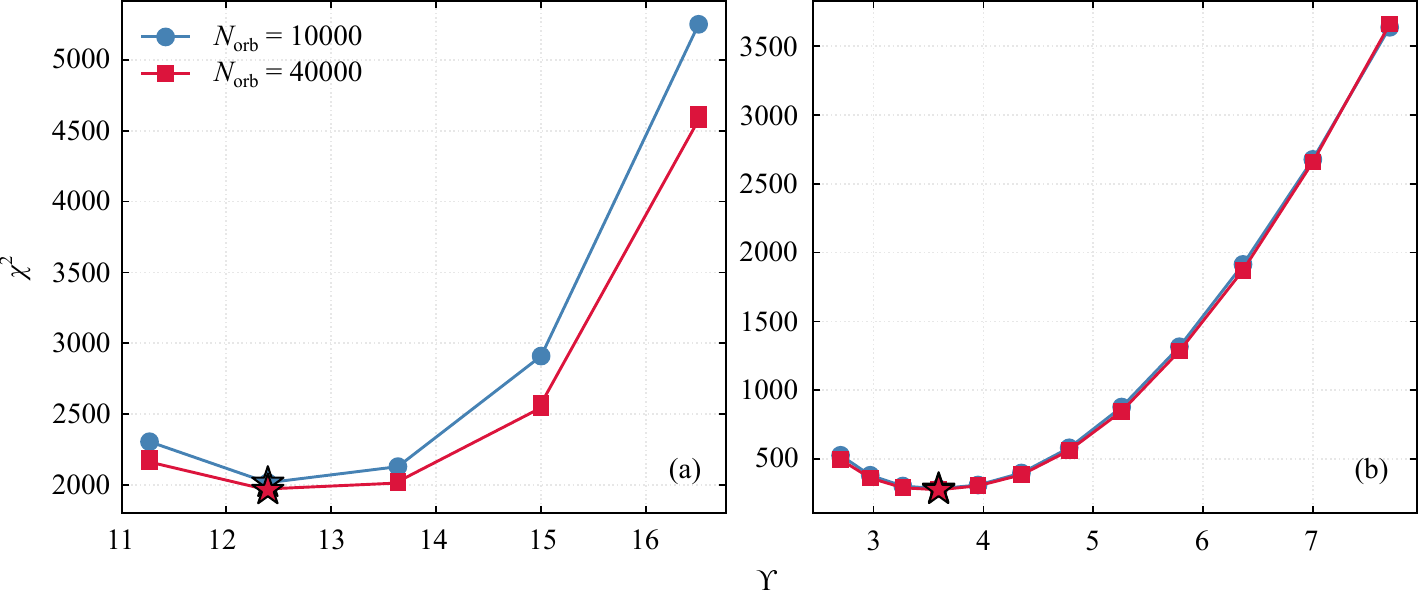}
\caption{Dependence of $\chi^2$ on the mass-to-light ratio $\Upsilon$ for fixed optimal values of $r_{\rm scale}$ and $v_{\rm circ}$ for PGC~35706 (a) and LEDA~2220522 (b). Blue circles and red squares show models with $N_{\rm orb}=10,000$ and 40,000 orbits, respectively. The asterisk marks the optimal $\Upsilon$ used in the final model. Agreement between the curves near the minimum confirms that 10,000 orbits are sufficient for locating the optimum.}
\label{fig:chi2ups}
\end{figure*}

The dependence of $\chi^2$ on $\Upsilon$ for the optimal fixed values of $r_{\rm scale}$ and $v_{\rm circ}$ is shown in Fig.~\ref{fig:chi2ups}. Because changing $\Upsilon$ only scales the velocities in the orbit library by $\sqrt{\Upsilon}$ and does not require the potential to be recalculated, this direction in parameter space can be treated almost independently of $(r_{\rm scale},v_{\rm circ})$, consistent with the three-level minimization described above. The minimum of $\chi^2(\Upsilon)$ is pronounced for both galaxies and corresponds to the values in Table~\ref{tab:model}. Figure~\ref{fig:chi2ups} also compares models with $N_{\rm orb}=10,000$ and 40,000. Near the optimum, the curves agree within the statistical $\chi^2$ fluctuations caused by the finite number of orbits ($1/\sqrt{N}$ per line-of-sight velocity bin), as expected when a discrete orbital ensemble converges to a continuous distribution function (Schwarzschild, 1979). The discrepancy far from the minimum at large $\Upsilon$ for PGC~35706 is explained by these fluctuations and does not affect the minimum. The search for optimal input parameters can therefore use 10,000 orbits, while the final model uses 40,000 to suppress noise in the component maps. A much smaller library ($N_{\rm orb}\sim10^3$) is unsuitable because the relative $\chi^2$ fluctuation reaches about 3\%, comparable to the width of local minima, preventing reliable identification of the global optimum.

\subsection{Schwarzschild Modeling}

The Schwarzschild method produces a dynamical model with the following output characteristics.
\begin{enumerate}
\item The orbital-weight distribution is a discrete approximation to the distribution function in the space of integrals of motion. The weights $w_i$ are determined by solving a quadratic-programming problem with non-negativity constraints and provide the best fit to the observations subject to dynamical self-consistency.
\item The internal kinematic structure includes three-dimensional mean-velocity fields and components of the velocity-dispersion tensor. These quantities are calculated by weighted summation of the contributions of individual orbits and allow the anisotropy of the velocity distribution to be studied.
\item The orbital composition is analyzed through the distribution of orbits over integrals of motion. For axisymmetric systems, the key parameters are the energy $E$ and the $z$ component of angular momentum $L_z$; triaxial systems use additional approximate integrals. Of particular interest is the circularity parameter $\lambda_z\equiv\overline{L_z}/L_{\rm circ}(E)$, which characterizes the degree of orbital elongation.
\item The gravitational-potential profile and mass distribution are fundamental modeling results. The method allows the stellar mass-to-light ratio $\Upsilon$ and dark-halo parameters to be estimated. These estimates may be degenerate, especially when the kinematic data have limited spatial coverage.
\end{enumerate}

Statistical indicators of model quality include $\chi^2$, which characterizes the agreement between the model predictions and the observations. Analyzing the dependence of $\chi^2$ on the potential parameters allows confidence intervals to be constructed, although rigorous statistical interpretation must account for the large number of hidden parameters represented by the orbit weights.

\section{Dynamical Modeling Results}

This section presents the main dynamical-modeling results for PGC~35706 and LEDA~2220522 and discusses their structural organization in terms of individual orbital components. The models allow the contributions of different subsystems to be separated quantitatively and their kinematic and morphological properties to be compared with those previously found in galaxies with counter-rotating disks and inconsistent kinematics (Schwarzschild, 1979; van den Bosch et al., 2008; Zhu et al., 2018a; Falc\'{o}n-Barroso and Martig, 2021; Bao et al., 2022, 2024).

\begin{figure*}[t]
\centering
\includegraphics[width=\textwidth]{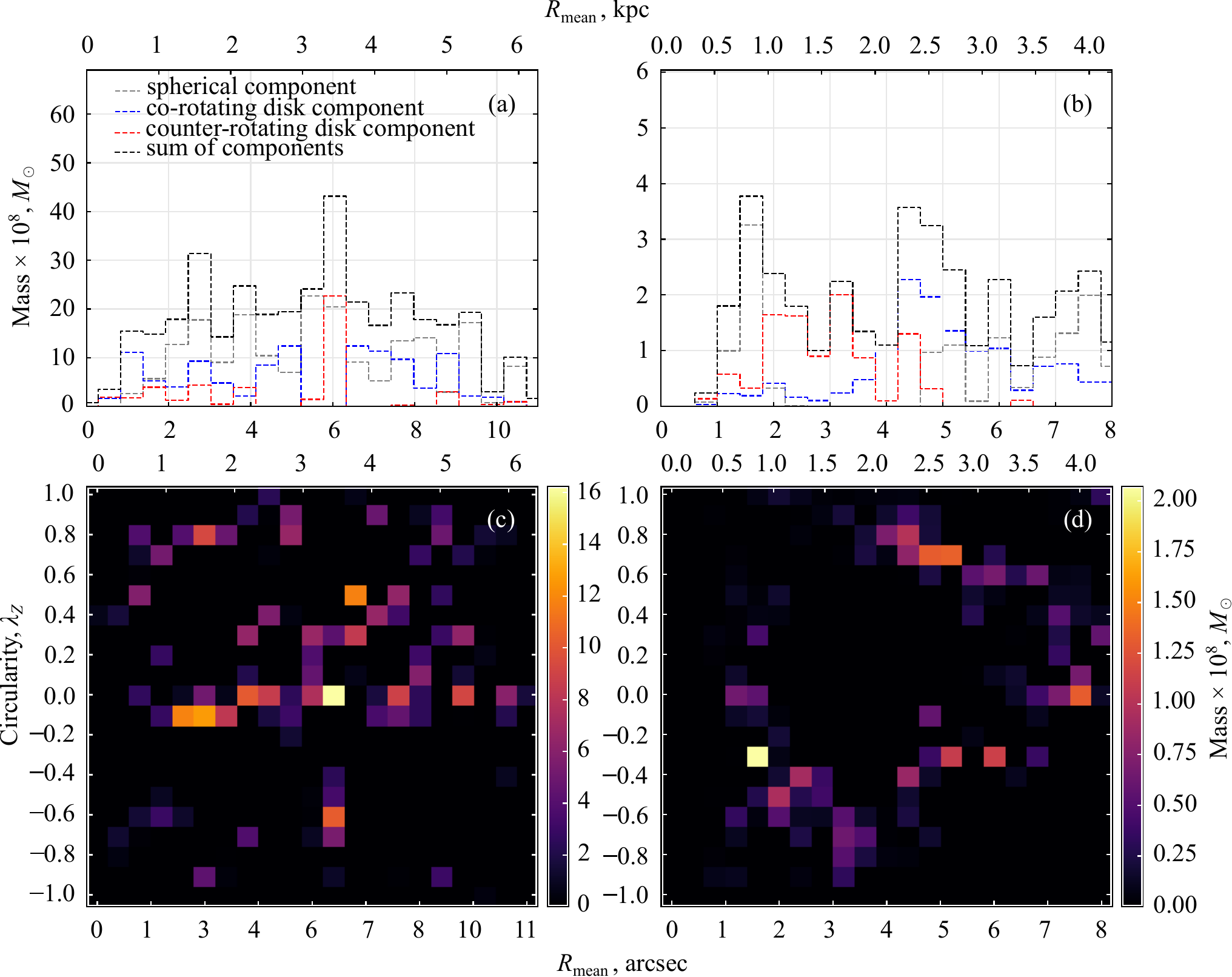}
\caption{Panel (a): component mass as a function of radius from the kinematic model of PGC~35706. Gray, blue, and red dashed lines show the radial mass distributions of the spheroidal component, co-rotating disk, and counter-rotating disk; the black dashed line is their sum. In panels (a) and (b), the lower abscissa is in arcseconds and the upper abscissa in kiloparsecs. Panel (c): orbital circularity $\lambda_z$ as a function of radius for PGC~35706. The spheroidal component has $0.35\gtrsim\lambda_z\gtrsim-0.35$, while the co-rotating and counter-rotating disks have $1\gtrsim\lambda_z\gtrsim0.35$ and $-0.35\gtrsim\lambda_z\gtrsim-1$, respectively. The apparent discreteness in panels (c) and (d) is due to the finite orbit library ($N_{\rm orb}=40,000$); the integrated component fractions and radial mass profiles are independent of it. Panels (b) and (d) show the same quantities for LEDA~2220522.}
\label{fig:masscirc}
\end{figure*}

\subsection{Structural Components of the Galaxies}

Dynamical models of both galaxies reveal several orbital components that can be interpreted naturally as a spheroidal component (bulge/halo) and two disk subsystems rotating in opposite directions. They are distinguished using the orbital circularity parameter $\lambda_z$, consistently with population-orbit modeling and analyses of counter-rotating disks (Zhu et al., 2018a,b, 2020; Bao et al., 2024). The total stellar mass of the model extrapolated beyond the MaNGA field of view (Table~\ref{tab:model}) is $M_{*,\rm tot}\approx6.2\times10^{10}\,M_\odot$ for PGC~35706 and $0.55\times10^{10}\,M_\odot$ for LEDA~2220522, close in order of magnitude to the photometric estimates of $4.7\times10^{10}\,M_\odot$ and $1.0\times10^{10}\,M_\odot$ from Wake et al. (2017). The discrepancy is less than a factor of two and is consistent with the systematic uncertainty associated with the dark-halo parametrization and the MGE approximation of the surface brightness.

\begin{figure*}[t]
\centering
\includegraphics[width=0.95\textwidth]{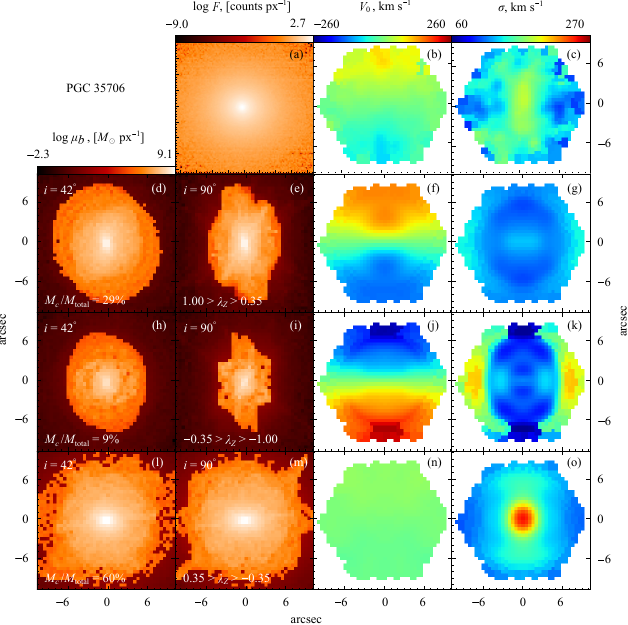}
\caption{Morphology and kinematics of the three components of PGC~35706. Panels (a)--(c) show the observed surface brightness, line-of-sight velocity, and velocity dispersion. Panels (d)--(o) show the surface density at inclination angles $i=90^\circ$ and $42^\circ$, line-of-sight velocities, and velocity dispersion for the two disk components and the spheroidal component (at $i=42^\circ$) obtained from the simulations. Each component corresponds to the indicated range of $\lambda_z$, and its fraction of the total mass is shown.}
\label{fig:pgccomp}
\end{figure*}

Figure~\ref{fig:masscirc} shows the radial mass profiles of the orbital components of PGC~35706 and the orbit distribution in the $R_{\rm mean}$--$\lambda_z$ plane. Most of the mass within $10''$ is concentrated in an almost non-rotating subsystem with $|\lambda_z|\lesssim0.35$, whose fraction is $M_c/M_{\rm total}\approx0.60$, where $M_c$ is the mass of the corresponding orbital component and $M_{\rm total}$ is the total mass of all model orbits within the MaNGA field of view. This subsystem is interpreted as a spheroid supported by stellar velocity dispersion. Two disks stand out against this background: the co-rotating disk ($1>\lambda_z>0.35$, $M_c/M_{\rm total}\approx0.29$) and the counter-rotating disk ($-1<\lambda_z<-0.35$, $M_c/M_{\rm total}\approx0.09$). The counter-rotating disk contributes only about 9\% of the total stellar mass and is concentrated at $R\sim3''$--$8''$. Its mass profile has a local maximum near $R\approx6''$, consistent with the ring-like structure in maps of the same component (Fig.~\ref{fig:pgccomp}); the stability of this feature is discussed below in connection with the choice of the $|\lambda_z|$ boundary.

\begin{figure*}[t]
\centering
\includegraphics[width=0.73\textwidth]{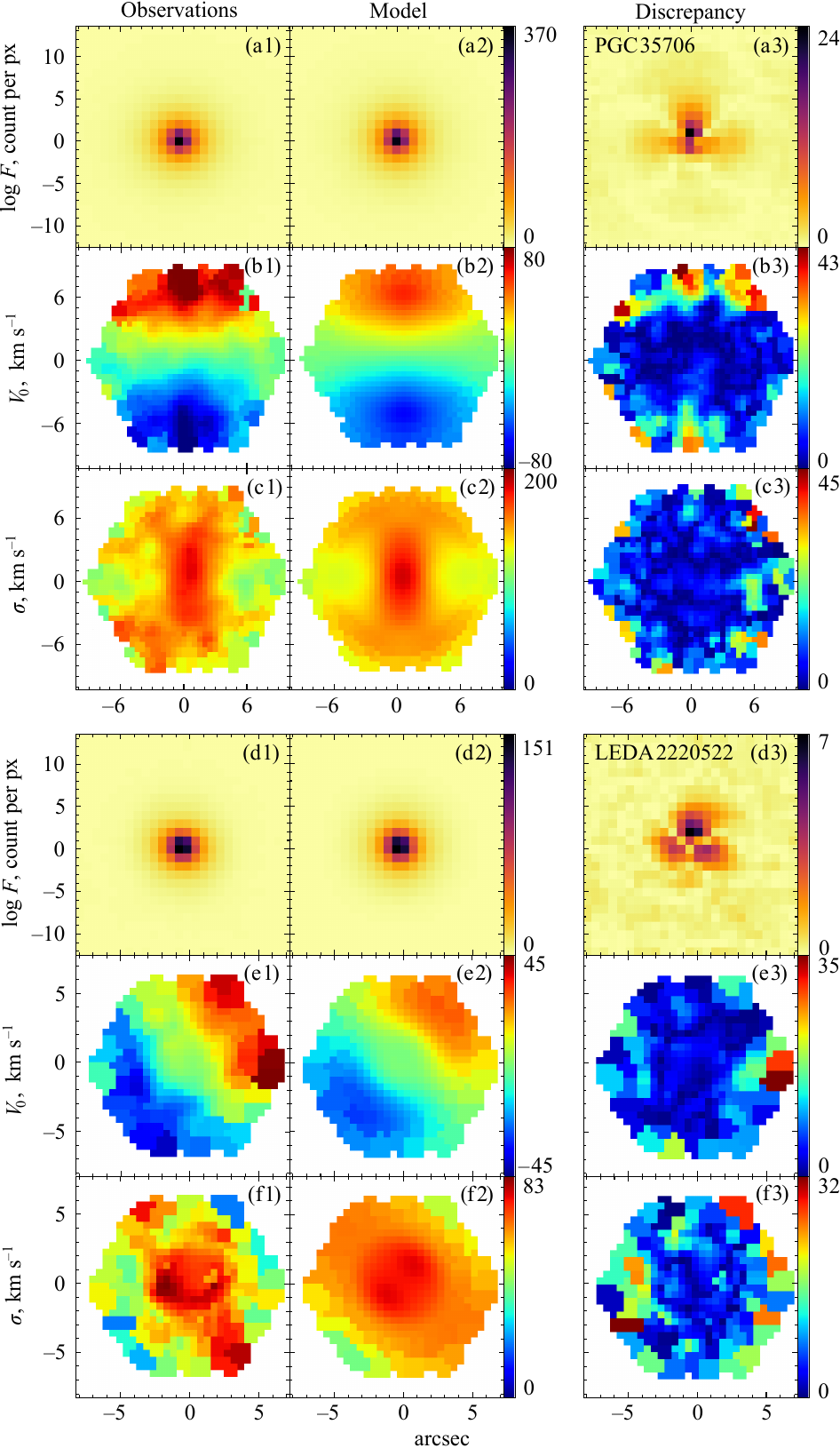}
\caption{Comparison of the photometry (panels (a) and (d)), line-of-sight velocity fields (panels (b) and (e)), and velocity dispersion (panels (c) and (f)) between observations and models of PGC~35706 and LEDA~2220522. The first column contains the observed data, the second the model, and the third the data-minus-model residuals.}
\label{fig:obsmodel}
\end{figure*}

The morphology and kinematics of the isolated components of PGC~35706 are shown in Fig.~\ref{fig:pgccomp}. The counter-rotating component contains a polar ring and disks inclined by approximately $30^\circ$ relative to the plane of the sky. Comparing the model line-of-sight velocity and velocity-dispersion fields with the observed MaNGA maps (Fig.~\ref{fig:obsmodel}) shows that the combination of the spheroidal and two disk subsystems reproduces the large-scale kinematics well, including the change in velocity sign along the major axis and the shape of the dispersion contours.

\begin{figure*}[t]
\centering
\includegraphics[width=0.95\textwidth]{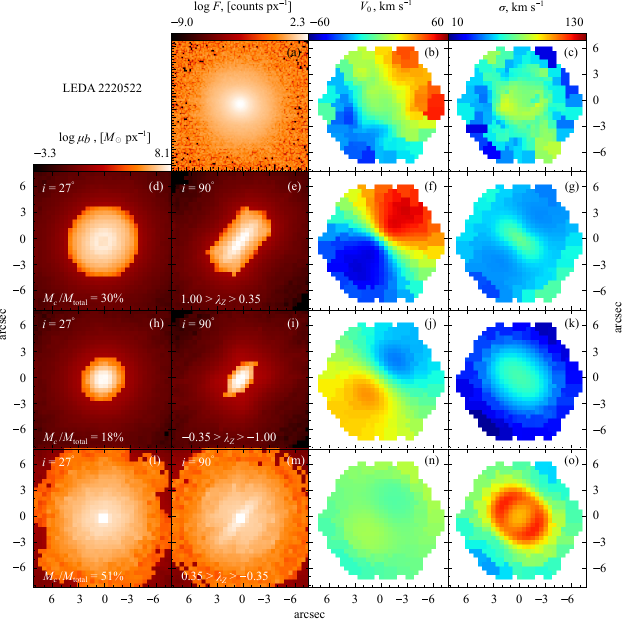}
\caption{Same as Fig.~\ref{fig:pgccomp}, but for LEDA~2220522.}
\label{fig:ledacomp}
\end{figure*}

For LEDA~2220522, similar mass and orbital-distribution profiles (Fig.~\ref{fig:masscirc}) show three components: a mass-dominant spheroid ($M_c/M_{\rm total}\approx0.51$), a co-rotating disk (about 0.30), and a counter-rotating disk (about 0.18). The co-rotating disk is thicker and more extended, dominating at $R\lesssim5''$, whereas the counter-rotating disk is thin and more compact, with most of its mass at $R\lesssim3.5''$. The co-rotating component also exhibits a polar ring inclined to the disk by approximately $60^\circ$, while the counter-rotating component contains a thin low-mass disk with $|\lambda_z|\approx0.5$--0.7. Figure~\ref{fig:ledacomp} shows that the disk subsystems rotate in opposite directions against the mass-dominant spheroid and that the model line-of-sight velocity and velocity-dispersion fields agree with the observed MaNGA maps (Fig.~\ref{fig:obsmodel}).

Comparing the orbital-component kinematics with the gas kinematics shows that the ionized-gas line-of-sight velocity field in both galaxies is closer to that of the counter-rotating disk. In PGC~35706, the model velocity map of the counter-rotating disk correlates well with the observed gas-velocity map (Figs.~\ref{fig:manga} and \ref{fig:pgccomp}). In LEDA~2220522, the gas-velocity contours follow the thin disk with $\lambda_z<0$ (Figs.~\ref{fig:manga} and \ref{fig:ledacomp}); the agreement between the dispersion maps is less diagnostic because the dispersion in this galaxy is largely determined by the massive spheroid. This is consistent with the interpretation of counter-rotating disks as having formed from external gas (Coccato et al., 2011, 2013; Katkov et al., 2016; Bao et al., 2022, 2024).

To test the stability of the decomposition with respect to the $|\lambda_z|$ boundary, component maps were also constructed for $|\lambda_z|=0.25$ and 0.45. When the boundary is shifted from 0.35 to 0.45, the fractions change only slightly: for PGC~35706, spheroid $0.60\to0.65$, co-rotating disk $0.29\to0.25$, and counter-rotating disk $0.09\to0.08$; for LEDA~2220522, $0.51\to0.58$, $0.30\to0.28$, and $0.18\to0.13$, respectively. This low sensitivity to tightening the boundary (less than five percent) is consistent with the $|\lambda_z|=0.35$ boundary lying outside the dense cloud of spheroidal orbits in the $(R,\lambda_z)$ diagrams (Fig.~\ref{fig:masscirc}). Shifting the boundary to $|\lambda_z|=0.25$ has a larger effect, especially in LEDA~2220522, where the spheroid fraction drops to 0.35 and the counter-rotating disk fraction increases to 0.28, because orbits naturally belonging to the spheroid then enter the disk components. The qualitative conclusions -- spheroid dominance in both galaxies, the small counter-rotating mass fraction in PGC~35706, and the polar-ring and inclined-disk morphology -- are preserved over the tested range $|\lambda_z|=0.25$--0.45.

\subsection{Possible Formation Scenarios}

The inferred orbital structure allows the formation pathways of the isolated components to be discussed in the context of current scenarios for the evolution of galaxies with inconsistent kinematics. Observational and numerical studies show that counter-rotating disks and complex multispin configurations are rarely explained solely by internal dynamical processes such as bar-driven evolution or global disk instabilities (Evans and Collett, 1994; Corsini, 2014). External processes play a substantial role: gas accretion from the intergalactic medium or cosmological filaments and minor mergers with gas-rich satellites (Bournaud et al., 2005; Katkov et al., 2013, 2016; Xu et al., 2022; Bao et al., 2022; Bevacqua et al., 2022).

For PGC~35706, the massive spheroidal component with $M_c/M_{\rm total}\approx0.60$, the stability provided by stellar velocity dispersion, and multiple inclined and polar disk structures in the co-rotating and counter-rotating components indicate a complex interaction history. The combination of a polar ring, inclined disks, and associated gas kinematics arises naturally through sequential accretion of gas and/or minor companions with different angular momenta (Katkov et al., 2016; Falc\'{o}n-Barroso and Martig, 2021; Bao et al., 2024). The agreement of the gas kinematics with the counter-rotating component (Figs.~\ref{fig:manga} and \ref{fig:pgccomp}) suggests that an accretion-formed gas disk played a key role in producing the corresponding orbital subsystem, as in counter-rotating disks in MaNGA and other integral-field spectroscopic surveys (Coccato et al., 2011, 2013; Jin et al., 2020; Santucci et al., 2022).

In LEDA~2220522, the non-rotating spheroid is the dominant mass component ($M_c/M_{\rm total}\approx0.51$), while the co-rotating and counter-rotating disks contribute approximately 0.30 and 0.18. The disk mass ratio, close to 0.6, makes LEDA~2220522 more similar to classical counter-rotating systems (Rix et al., 1992; Rubin et al., 1992; Bao et al., 2022) than PGC~35706, whose counter-rotating disk accounts for only 9\% of the total mass; nevertheless, the spheroid in LEDA~2220522 is more massive than either disk component. The thick co-rotating disk, dominant at radii up to $5''$, can be interpreted as a primordial disk formed early in the galaxy's evolution. The thin counter-rotating disk, with most of its mass at $R\lesssim3.5''$, is a product of later gas accretion with angular momentum opposite to the pre-existing stellar subsystem (Coccato et al., 2011; Johnston et al., 2013; Bao et al., 2024). The thin low-mass disk in the counter-rotating component with $|\lambda_z|\approx0.5$--0.7 has a lower radial velocity dispersion (Fig.~\ref{fig:ledacomp}), which may indicate recent formation. As in Bao et al. (2024), where the secondary disk is less massive and more compact, the LEDA~2220522 model has a counter-rotating component with a smaller characteristic radius and somewhat lower mass than the co-rotating disk (Figs.~\ref{fig:masscirc} and \ref{fig:ledacomp}), perhaps indicating a relatively recent, mass-limited gas-accretion episode.

Deep DESI images show no prominent signatures of a recent major merger, such as bright tails, shells, or strong distortions (Ji et al., 2014; Dey et al., 2019; Li et al., 2021). This is consistent with numerical models in which gas exchange with the environment and minor and intermediate-mass-ratio mergers with gas-rich companions can form counter-rotating disks without destroying the global structure of the galaxy (Bournaud et al., 2005; Lagos et al., 2015). The morphology, stellar and gas kinematics, and orbital structure of PGC~35706 and LEDA~2220522 together indicate that long-term gas accretion and/or a series of minor mergers with gas-rich companions are the most likely mechanisms for forming the detected counter-rotating components.

A promising direction for future work is a quantitative comparison of the ages and metallicities of the stellar subsystems obtained from spectral analysis with the orbital components of the model, as in population-orbit studies (Zhu et al., 2020). This will allow more precise dating of accretion episodes and refinement of the evolutionary scenarios for counter-rotating disks in each galaxy.

\clearpage
\section{Conclusions}

We have presented dynamical models of the two galaxies with inconsistent kinematics, PGC~35706 and LEDA~2220522, constructed with the Schwarzschild orbit-superposition method in an axisymmetric approximation from MaNGA integral-field spectroscopy and deep DESI photometry. The main results are as follows.

\begin{enumerate}
\item \textit{Orbital decomposition.} In both galaxies, the dynamical models separate the stellar system into three main orbital components:
  \begin{enumerate}
  \item a dispersion-supported spheroidal component (bulge/halo) with low circularity, $|\lambda_z|\lesssim0.35$, accounting for about 60\% of the mass within the MaNGA field of view in PGC~35706 and about 51\% in LEDA~2220522;
  \item a co-rotating disk ($0.35<\lambda_z<1$), an older and more extended component. In PGC~35706 it accounts for about 29\% of the mass. In LEDA~2220522 it is thick, dominates at $R\lesssim5''$, and contains a ring inclined by about $60^\circ$ to the disk plane;
  \item a counter-rotating disk ($-1<\lambda_z<-0.35$), a less massive and more compact secondary disk. In LEDA~2220522 it is thin and concentrated at $R\lesssim3.5''$; in PGC~35706 it has a complex structure including features of a polar ring and tilted disks.
  \end{enumerate}

\item \textit{Agreement with the observed kinematics.} The models satisfactorily reproduce the observed line-of-sight velocity fields and stellar velocity dispersions from MaNGA. An important result is the agreement between the ionized-gas kinematics traced by H$\alpha$ and the counter-rotating stellar disk in both galaxies. This indicates a common origin of the gas and stars of the secondary disk in externally accreted gas whose angular momentum is opposite to that of the pre-existing disk.

\item \textit{Formation scenarios.} The inferred orbital structure supports the formation of counter-rotating disks through external gas accretion. For LEDA~2220522, the comparable masses of the disk components (counter-rotating to co-rotating mass ratio $\approx0.60$), the compact secondary disk, and the polar-ring morphology indicate several relatively recent gas-accretion episodes, probably from the intergalactic medium or through a minor merger with a gas-rich companion. In PGC~35706, the still more complex morphology and the spheroid dominance may reflect a longer, multistage accretion history. The ages and chemical compositions of the structures must nevertheless be clarified before unambiguous conclusions are drawn about the dynamical histories of the galaxies.

\item \textit{Method validation.} The results demonstrate the effectiveness of the Schwarzschild method implemented in AGAMA and \texttt{Forstand} for analyzing complex galactic kinematics. The circularity parameter $\lambda_z$ is an informative component-separation criterion consistent with modern population-orbit studies.
\end{enumerate}

Future work should combine dynamical modeling with stellar-population analysis. Comparing age and metallicity maps from full spectral fitting with the orbital components will enable direct dating of disk-formation episodes and refinement of accretion scenarios. Nonparametric reconstruction of the LOSVD can help separate nearby kinematic components more accurately in galaxy centers. Extending the analysis to a larger sample of galaxies with inconsistent kinematics from MaNGA and DESI will permit a statistical assessment of the frequency and properties of such systems and hence of the role of external accretion in galaxy evolution in the local Universe.

Thus, dynamical modeling with the Schwarzschild method is a powerful tool for decomposing galaxies into orbital components and studying their formation histories, helping to clarify the role of external processes -- gas accretion and minor mergers -- in producing complex galactic kinematics.

\section*{Funding}
The work was supported by the Russian Science Foundation, grant 22-12-00080.

\section*{Conflict of Interest}
The authors declare no conflict of interest.

\makeatletter
\let\originalbibitem\bibitem
\renewcommand{\bibitem}{\@ifnextchar[\numberedbibitem\originalbibitem}
\def\numberedbibitem[#1]#2{\originalbibitem{#2}}
\makeatother

\end{document}